\PassOptionsToPackage{dvipsnames}{xcolor}
\documentclass[conference]{IEEEtran}
\IEEEoverridecommandlockouts
\usepackage[T1]{fontenc}
\usepackage{cite}
\usepackage{amsmath,amssymb,amsfonts}
\usepackage{algorithm}
\usepackage{algorithmic}
\usepackage{array}
\usepackage{booktabs}
\usepackage{graphicx}
\usepackage{textcomp}
\usepackage{xcolor}
\usepackage{siunitx}
\DeclareSIUnit{\request}{req}
\DeclareSIUnit{\token}{tok}
\usepackage{layouts}
\usepackage{listings}
\usepackage{nicefrac}
\usepackage{balance}
\usepackage{nowidow}

\usepackage[hidelinks]{hyperref}
\usepackage[capitalise]{cleveref}

\definecolor{darkblue}{rgb}{0.0, 0.0, 0.55}
\newcommand{\confname}[1]{\textsf{#1}}
\newcommand{\sNaive}{\confname{naive}}
\newcommand{\sTn}{\confname{tn}}
\newcommand{\sLitellm}{\confname{litellm}}
\newcommand{\sBp}{\confname{bp}}

\begin{document}
\bstctlcite{IEEEexample:BSTcontrol}

\title{Dispatch Control for Priority-Based, Multi-Tenant Workflows under Multi-Resource External Rate Limits
    \thanks{This work was conducted in collaboration with Forgent AI GmbH, which provided the use case and production insights. S.~Kosorin and L.~Wossnig are also affiliated with Forgent AI GmbH.}
}

\author{\IEEEauthorblockN{Stanislav Kosorin\IEEEauthorrefmark{1}, Trever Schirmer\IEEEauthorrefmark{1}, Leonard Wossnig\IEEEauthorrefmark{2}, David Bermbach\IEEEauthorrefmark{1}}
    \IEEEauthorblockA{\IEEEauthorrefmark{1}\textit{Technische Universit\"at Berlin, Scalable Software Systems Research Group} \\
        \{stk,ts,db\}@3s.tu-berlin.de}
    \IEEEauthorblockA{\IEEEauthorrefmark{2}\textit{University College London, Department of Computer Science}\\
        leonard.wossnig@forgent.ai}
}

\maketitle
\begin{abstract}
    Workflow engines orchestrate pipelines of activities, which often call external APIs, on a fixed pool of worker slots, the threads that run them.
    When an API enforces a hard rate limit shared by all tenants and priorities, a throttled call that retries after a backoff keeps its worker slot and starves other tenants.
    Large Language Model provider APIs are particularly complex:
    their per-organization caps are low and bind independently on requests \emph{and} tokens per minute.
    The operator can neither reconfigure the cap nor observe saturation before a request fails, so classical fair-queuing assumptions do not hold.
    We present \emph{Backpressure}, a worker-side admission gate.
    It checks each call's cost atomically against shared admission state and enforces fairness over requests and tokens.
    A denied call releases its worker slot and waits on a durable timer, so waiting no longer blocks execution capacity.
    We compare against a naive retry, the AI gateway LiteLLM, and the built-in throttle of the workflow engine Temporal.
    On a workload with three priority classes, Backpressure raises fairness (Jain's index over each queue's fraction of its entitled share) from \num[mode=match, reset-text-series=false, text-series-to-math=true]{0.60} to \num[mode=match, reset-text-series=false, text-series-to-math=true]{0.89}, where \num[mode=match, reset-text-series=false, text-series-to-math=true]{1.0} is ideal.
    It also cuts high-priority tail latency roughly \qty[mode=match, reset-text-series=false, text-series-to-math=true]{8}{\times}.
    When tokens are the binding limit, all request-based baselines serve every priority equally, whereas Backpressure keeps the priority order and reaches a dominant-resource fairness of \num[mode=match, reset-text-series=false, text-series-to-math=true]{\sim 0.99}.
    Admission costs one round trip to shared state, adds no measurable latency when the cap does not bind, and fails open if shared state is unavailable.
\end{abstract}


\section{Introduction}
\label{sec:introduction}

Workflow engines run production pipelines for many tenants as sequences of steps, called activities.
Each running activity occupies one of a fixed number of worker slots, the engine's scarce resource.
Activities routinely call rate-limited external APIs, and many tenants and priority classes share one API ceiling.
Noisy neighbors can exhaust it, but the operator does not own it and so cannot enforce fairness directly.
Large Language Model (LLM) providers are not the only source of such caps, but they are a particularly hard case.
Their per-organization caps on Requests Per Minute (RPM) and Tokens Per Minute (TPM)~\cite{openai_rate_limits,anthropic_rate_limits,azure_openai_quotas} are orders of magnitude below traditional cloud services.
The two also bind independently, so the request mix, not request count alone, decides which one saturates.
Yet workflow engines have no mechanism to share such a ceiling among tenants.
They dispatch by queue state and worker availability and learn of overload only when an \texttt{HTTP 429} arrives.
This leaves operators three workarounds, none of which decides, before a call is sent, which tenant may use the shared cap.
(a)~In-process SDK retries couple worker concurrency to provider backoff.
(b)~Engine-native throttles limit dispatch without observing upstream 429s.
(c)~HTTP retry gateways absorb retries while holding the caller's connection open~\cite{litellm,portkey,helicone,kong_ai}.
In every case, a noisy tenant can starve low-volume tenants.
Because rejected calls also count against the external cap, the retries turn into a self-inflicted denial of service.

Fair-share scheduling itself is well understood, e.g., Weighted Fair Queuing (WFQ)~\cite{demers1989wfq} and Deficit Round Robin (DRR)~\cite{shreedhar1995drr}, but these algorithms assume that the operator controls the contended resource.
Here the cap belongs to the provider and cannot be reconfigured.
We present \emph{Backpressure}, an admission control approach for workflow engines that acts only on the operator's own workers.
A denied request releases its worker slot and defers on a durable timer, so the wait costs no execution capacity.
Our prototype is a plugin for Temporal~\cite{temporal}.
Before an activity issues its first HTTP call, the worker checks the call's cost atomically against shared admission state.
That state enforces the global cap, a per-priority split, and a weighted fair share per tenant within a priority.
A denial returns a precise retry interval for the workflow's durable sleep.
On a cross-priority workload, Backpressure raises fairness from 0.60 to 0.89 (achievement-Jain~\cite{jain1998quantitativemeasurefairnessdiscrimination}, 1.0 is an even split).
It also cuts high-priority tail latency \qty{8}{\times} and drops the 429 share from \qty{23}{\percent} to \qty{4}{\percent}, at \qty{\sim 10}{\percent} lower mean throughput.
When tokens bind, every request-only baseline flattens priority, while Backpressure keeps the configured ordering at \num{\sim 0.99} dominant-resource fairness (DRF).
Our main contributions are:
\begin{itemize}
    \item We formalize fair sharing of hard, externally enforced, multi-resource rate limits in multi-tenant workflow engines (\cref{sec:background,sec:problem}).
    \item We design Backpressure, a worker-side gate that admits each call against shared request and token buckets with priority and dominant-resource fairness, and turns a denial into a durable timer that frees the worker slot (\cref{sec:design,sec:timer}).
    \item We release Backpressure as open source and evaluate it against three baselines with a mock rate-limited LLM API (\cref{sec:evaluation}).
\end{itemize}

\section{Background}
\label{sec:background}
A workflow engine runs multi-step applications on behalf of many tenants.
Each step is an activity, dispatched from a task queue to a pool of workers, and occupies a worker slot while it runs.
Engines add durable state, at-least-once execution, and deterministic replay, so application code does not re-implement fault tolerance~\cite{temporal,stepfunctions,durable_functions,airflow}.
They dispatch by queue depth and availability and expose only static priorities and per-queue rate limits as admission controls.
Recent work on engine internals improves durability and throughput but keeps this dispatch model~\cite{zhuang2023exoflow,burckhardt2022netherite,jia2021boki}.
Downstream services are assumed elastic, and the worker slot is the only limited resource the engine manages.
External rate-limited providers break that assumption.
Their per-organization ceilings~\cite{openai_rate_limits,anthropic_rate_limits,azure_openai_quotas} are hard limits outside the operator's control, and many other providers impose similar ones:
payment processors cap \emph{account} transactions per second, and SaaS platforms cap query volume.
The engine can neither see these global counters nor reserve capacity against them.
Overload therefore surfaces only as \texttt{HTTP 429 Too Many Requests}, often with a \texttt{Retry-After} hint, and only after a slot is committed.

\subsection{Three failure modes under contention}

\begin{figure}[t]
    \centering
    \includegraphics[width=\columnwidth]{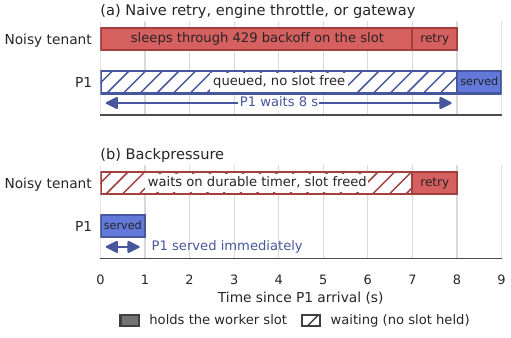}
    \caption{One worker slot under overload. A high-priority request (P1) arrives at $t{=}0$ as a noisy tenant enters a \qty{7}{\second} retry backoff. In (a), the solid bar holds the slot. In (b), the deny frees the slot at once, indicated by a hatched bar.}
    \label{fig:slot-trap}
\end{figure}

When the cap binds, the three common responses fail differently but with the same consequence.

\paragraph{Mode A: Oblivious SDK retries}
The simplest approach wraps every external call in the provider SDK's or the engine's activity-level retry policy.
Both treat 429 as transient, sleep on the worker thread, and retry in place.
The slot stays occupied for the entire backoff, and a single noisy tenant can rapidly deplete the worker slots.

\paragraph{Mode B: Engine-native throttles}
Temporal and AWS Step Functions expose per-queue Requests Per Second (RPS) limits, priority keys, and fairness keys.
These throttle how fast the engine dispatches work, not how fast it succeeds upstream.
Once the provider cap saturates, every dispatched activity returns 429.
Re-enqueuing that work at the same priority amplifies the offered load at the cap, the trigger pattern of metastable failure~\cite{metastable}.
The throttle is also request-only, so when TPM binds it never engages and behaves like Mode~A.

\paragraph{Mode C: HTTP retry gateways}
Gateways such as LiteLLM, Portkey, Helicone, and Kong AI Gateway~\cite{litellm,portkey,helicone,kong_ai} are designed to hide retries from application code.
They hold the caller's TCP connection open for the full \texttt{Retry-After} interval.
To the engine, the activity remains in-flight and holds its slot, as in Mode~A.
Across modes, a tenant that ignores the cap consumes worker concurrency in proportion to its retry budget rather than its admitted throughput (\cref{fig:slot-trap}).
Low-volume tenants therefore starve for slots even when their own load alone would never be throttled.

\section{Problem Formalization}
\label{sec:problem}

These failures lead to four requirements that no existing layer meets together.
(a)~Global visibility against the externally enforced cap, not per-worker estimates.
(b)~Metering of requests and tokens that stays fair whichever dimension binds.
(c)~Fair share across tenants and priorities, not first-come-first-served dispatch.
(d)~Slot release on deny before the retry interval elapses, which decouples cluster concurrency from provider backoff.
We formalize this cooperative-sharing gap as an admission-control problem over a shared, multi-resource external cap.

\subsection{Setting}

Consider a workflow cluster serving tenants $\mathcal{T} = \{t_1, \ldots, t_N\}$ and priority classes $\mathcal{K} = \{k_1, \ldots, k_M\}$ with total order $k_1 \succ k_2 \succ \cdots \succ k_M$.
All activities share one external target that enforces resource dimensions $\mathcal{D}$, each with a hard ceiling of $C_d$ admissions of resource $d$ per unit time.
For an LLM provider, $\mathcal{D} = \{\text{RPM}, \text{TPM}\}$.
Each candidate activity $a$ carries a cost vector $c(a) = (c_d(a))_{d \in \mathcal{D}}$: one request and its (estimated) token count.
The operator does not own any $C_d$ and cannot raise it on demand.
It observes congestion only through rejected admissions (429s), which do not report which dimension bound.
Tenant $t$ has weight $w_t > 0$, priority $k$ weight $\omega_k > 0$, and queue $(t,k)$ is entitled in proportion to $w_t \omega_k$.
The engine has $S$ concurrent worker slots.
We assume:

\noindent\textbf{A1 (Static, known caps).} Every $C_d$ is fixed and set in the controller.
Bucket and share parameters derive from it, so a cap changing without notice invalidates them.
\textbf{A2 (Sole consumer).} The cluster alone uses the cap.
Outside traffic skips shared state and silently erodes the cap.
\textbf{A3 (Single cap scope).} One ceiling set covers the target.
Provider sub-quotas (e.g., per model or deployment) each need a bucket set.

\subsection{Admission decisions and required properties}

At each dispatch tick, an admission rule makes one of three decisions for every candidate activity $a$.
It may \textbf{admit} $a$ and issue the external call, or \textbf{defer} it for interval $\Delta_a$ while immediately returning the worker slot.
It may \textbf{fail} $a$ on terminal errors only (never as flow control).
We require five properties of any satisfactory rule:

\noindent\textbf{P1 (Cap respect).} For every dimension $d$, aggregate admitted cost $\sum_a c_d(a)$ must not exceed $C_d$ in any time unit of dimension $d$.

\noindent\textbf{P2 (Dominant-resource fair share).} Under sustained backlog, each queue $(t,k)$ is equalized on the share of its dominant resource $\arg\max_d c_d/C_d$, weighted by $w_t \omega_k$.
This is DRF~\cite{ghodsi2011drf} applied to admissions.
It reduces to WFQ~\cite{demers1989wfq,shreedhar1995drr} when one dimension binds for all tenants.

\noindent\textbf{P3 (Work conservation).} The binding dimension's cap must not idle while any queue is backlogged (a non-binding dimension necessarily idles once another saturates first).
Lower-priority queues consume an empty high-priority queue's share.

\noindent\textbf{P4 (Estimate reconciliation).} A cost known only as an estimate (e.g., tokens before the reply) must later be reconciled with its true value.
Otherwise its bucket drifts from $C_d$.

\noindent\textbf{P5 (Slot release on deny).} A deferred activity must free its slot before $\Delta_a$ elapses (requirement~(d)).

\subsection{Fairness-first scheduling}

P2 and P3 leave open how priority classes and tenant fairness interact.
Strict priority (drain $k_1$ before $k_2$, then $k_3$) is fragile under hard caps.
A persistent $k_1$ backlog from one tenant can monopolize the cap and starve every lower class.
Yet pure fair share without priority cannot meet tiered latency targets.
Backpressure thus adopts a \emph{fairness-first} policy.
The per-priority share (L2, \cref{sec:design:buckets}) is work-conserving, so $k_1$ may borrow idle capacity from $k_2$ but not starve it when both are backlogged.
Tenants within a class get weighted DRF share (L3).

\subsection{Scope}

We address dispatch control only for deployments that satisfy A1 to A3.
Beyond a worker-side wrapper, we change nothing in the engine, activity retry policy, or workflow code.
The allocator itself descends from WFQ~\cite{demers1989wfq}, DRR~\cite{shreedhar1995drr}, and DRF~\cite{ghodsi2011drf}.
What is new is where admission happens: inside the worker, on atomically updated shared state, with denial turned into a durable-timer deferral.

\section{Design}
\label{sec:design}
\begin{figure}[t]
    \centering
    \includegraphics[width=\columnwidth]{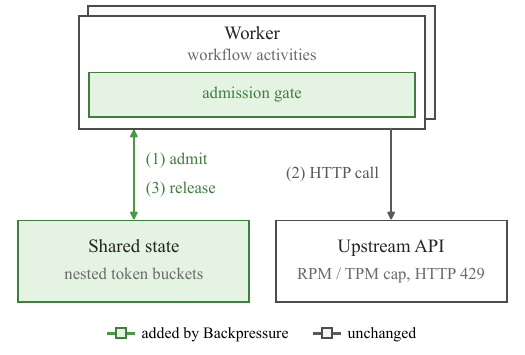}
    \caption{Admission path of one worker (all run the same gate against one shared state). (1)~One atomic operation on the shared buckets returns \texttt{admit} or \texttt{defer}$(\Delta)$. (2)~Only after a grant does the worker itself call the API, so admission state never sits on the call path. (3)~Once the provider reports usage, a signed release reconciles the token estimate. Green: added by Backpressure. Gray: unchanged (engine, workers, provider traffic).}
    \label{fig:architecture}
\end{figure}

Backpressure implements the admission rule of \cref{sec:problem} in the worker, directly before each rate-limited call, and leaves the engine unchanged (\cref{fig:architecture}).

\subsection{Admission at the call site}

The gate takes the tenant $t$, the priority $k$, and the call's cost vector $v=c(a)$.
It returns admit, or defer with interval $\Delta$.
On defer it does not wait in place but frees its slot and re-enters after $\Delta$ on a durable timer.
The gate thus needs nothing from the provider, only a durable engine sleep, which most engines offer~\cite{temporal,stepfunctions,durable_functions}.

\subsection{Nested token buckets}
\label{sec:design:buckets}

Each admission consults three nested layers that jointly enforce cap respect, work-conserving priority, and dominant-resource tenant fairness in one decision.

\noindent\textbf{L1 (Global cap).}
One bucket per dimension $d$, refilled at rate $C_d$.
Admitting $a$ consumes $c_d(a)$ from each, a hard ceiling every tenant and priority must clear.

\noindent\textbf{L2 (Per-priority share).}
One bucket per class $k$ and dimension $d$, nominally limited to $\omega_k C_d / \sum \omega$.
Work-conserving borrowing passes idle high-priority capacity to the next class in the same tick.
L2 is thus a soft yield, not a strict filter.

\noindent\textbf{L3 (Dominant-resource fair share).}
A single share bucket per active $(t,k)$ pair, measured in normalized dominant-resource units $\sigma(a)=\max_d c_d(a)/C_d$.
It refills at $\tfrac{w_t}{W_k}\,X^\star$ share units per second.
Here $W_k$ is the live weight sum at priority $k$, and $X^\star$ is the DRF equalization rate~\cite{ghodsi2011drf}
\begin{equation}
    X^\star = \min_d \frac{C_d\, W_k}{D_{k,d}},
    \qquad D_{k,d} = \sum_{(t,k)\ \text{active}} w_t\, \frac{c_d(t)}{\sigma(t)},
    \label{eq:xstar}
\end{equation}
with $c_d(t)$ tenant $t$'s current per-request cost and $\sigma(t)$ its dominant share.
$D_{k,d}$ is demand per unit of dominant share, so this refill drives the binding dimension (the arg\,min) to its cap while keeping weighted dominant shares equal.
The weight $\omega_k$ enters through the L2 shares rather than the L3 rate, so P2 holds within each class.
The nested refill is a fluid approximation of DRF, validated empirically (\cref{sec:eval:multi}).

An admission for $(t,k)$ succeeds only when every L1 and L2 dimension and the L3 share bucket hold enough tokens.
All are then debited atomically.
Otherwise, the rule returns $\Delta$, the earliest refill time among the depleted buckets.
$\Delta$ carries no jitter: synchronized wake-ups are harmless because admission calls are serialized and take milliseconds (\cref{sec:eval:overhead}), so the first grant debits the bucket and the others are deferred again.
Reducing L3 to one shared dominant-share bucket (rather than one weighted bucket per dimension) equalizes dominant share instead of per-resource share.
A token-heavy and a request-heavy tenant thus receive their fair split even though they stress different limits.

\subsection{Estimate reconciliation and reservation}
\label{sec:design:reconcile}

Before the reply, the token component of $c(a)$ is only an input-based estimate.
Backpressure debits it at admission and, once the provider reports usage, issues a signed release that debits or refunds the difference (P4).
This keeps the TPM bucket aligned with true consumption.
A gate that debited only the estimate would over-admit whenever completions run long.
Requests above the cap are rejected with 429, and on major providers these failed requests still count against the per-minute limit~\cite{openai_rate_limits}.
A transient burst above the cap therefore wastes rate budget rather than being free.
For such workloads, an optional \emph{reservation} mode debits the completion upper bound (\texttt{input}\,+\,\texttt{max\_tokens}) up front and refunds the unused share after the call.
This trades a little steady-state throughput for much less overshoot.

\subsection{Atomic admission on shared state}
\label{sec:design:atomic}

All buckets live in state shared by the worker pool, and one admission is one operation on it (\cref{alg:admit}).
The approach asks only that the store serialize this read-modify-write across workers.
It needs no lock or consensus round and costs one round trip of $O(|\mathcal{T}_k|)$ work.
This is the only mutable state Backpressure adds, and the reconciling release is a second, smaller operation of the same kind.
Releases are awaited but best-effort.
If a worker dies before its release lands, the gate errs conservatively with over-debited buckets and never over-admits (invariant I1 below).
Refill erases the residue within a burst window.

\begin{algorithm}[tb]
    \caption{Atomic multi-resource admission.}
    \label{alg:admit}
    \small
    \begin{algorithmic}[1]
        \REQUIRE tenant $t$, priority $k$, cost $c$, now $\tau$
        \ENSURE \texttt{ADMIT} or \texttt{DEFER}$(\Delta)$
        \STATE refill L1$_d$, L2$_{k,d}$ for all $d$, and L3$_{t,k}$ by elapsed time
        \STATE evict idle $(\cdot,k)$ entries, then scan peers for $W_k$ and $X^\star$
        \STATE $\sigma \gets \max_d c_d / C_d$, update L3$_{t,k}$ rate $\gets (w_t/W_k)\,X^\star$
        \IF{$\forall d:\ \mathrm{L1}_d \ge c_d \wedge \mathrm{L2}_{k,d}\ \text{ok}$ \textbf{and} $\mathrm{L3}_{t,k} \ge \sigma$}
        \STATE debit all buckets by $c$ (L3 by $\sigma$)
        \STATE \textbf{return} \texttt{ADMIT}
        \ELSE
        \STATE $\Delta \gets$ earliest refill over the depleted buckets
        \STATE \textbf{return} \texttt{DEFER}$(\Delta)$
        \ENDIF
    \end{algorithmic}
\end{algorithm}

\subsection{Invariants}

Admission must be linearizable, which confines shared state to one authoritative copy.
The approach does not replicate this copy; if it is lost, the gate fails open (I3) instead of hiding the failure.
Three operational invariants follow.

\noindent\textbf{I1 (No double counting).} Admission atomically checks and debits all dimensions, so concurrent workers cannot over-commit any cap.
\textbf{I2 (No worker-local state).} Workers hold no bucket state.
Weights and caps are read from shared state at decision time and may be hot-updated.
\textbf{I3 (Degraded open).} The gate fails open (\cref{sec:timer}), which preserves availability at the cost of temporary fairness loss.

\subsection{Prototype}
\label{sec:design:impl}

The evaluated prototype is a Python package for Temporal workers, which we publish as open source software\footnote{\url{https://github.com/3s-rg/backpressure}}.
Developers wrap external calls in a context manager:
\begin{lstlisting}
async with bp.admit(tenant=t, priority=k, cost=v):
    response = await provider.call(...)
\end{lstlisting}
\noindent The shared state is one Valkey primary~\cite{valkey}.
On defer the wrapper raises \texttt{BackpressureRetry} carrying $\Delta$, which the workflow turns into \texttt{workflow.sleep}$(\Delta)$.
A deployment adds a thin client to each worker pod and one Valkey node.

\section{Timer-Based Deferral}
\label{sec:timer}

The admission buckets decide when to admit work.
Timer-based deferral decides where the wait happens, and that choice separates Backpressure from the alternatives.

\subsection{Slot behavior during the wait}

As \cref{tab:slot-comparison} shows, all three modes of \cref{sec:background} hold the slot during a denial.
Backpressure releases it, as the wait is managed by the workflow runtime, not the worker thread.

\begin{table}[t]
    \centering
    \footnotesize
    \caption{Worker-slot behavior during the wait.}
    \label{tab:slot-comparison}
    \begin{tabular}{@{}lcc@{}}
        \toprule
        \textbf{Strategy}                  & \textbf{Slot freed} & \textbf{Cap-aware} \\
        \midrule
        Naive SDK retry (Mode~A)           & $\times$                & $\times$            \\
        Engine throttle (Mode~B)           & $\times$                & requests only       \\
        HTTP retry gateway (Mode~C)        & $\times$                & partial             \\
        LiteLLM proxy (client-side)        & $\times$                & partial             \\
        \textbf{Backpressure (ours)}       & $\checkmark$            & \textbf{both resources}  \\
        \bottomrule
    \end{tabular}
\end{table}

HTTP gateways cannot release slots because they sit on the call path after the workflow has committed a slot.
Even the LiteLLM proxy's client-side limiter, which pre-empts some overshoot, leaves the activity in-flight and its slot held.
Backpressure operates one layer earlier and translates a denial into a durable timer before any HTTP traffic is sent.

\subsection{Deferral as a durable timer}

On defer, the gate hands the wait to the engine: the workflow records the wake time in durable history and sleeps for $\Delta$.
Unlike a worker-side \texttt{sleep}, this frees the slot in the same dispatch tick.
When the timer fires, the engine re-schedules the activity, possibly on another worker, which runs admission again.
High-priority work thus flows through a fleet whose concurrency $S$ is no longer pinned by low-priority backoff.
The L3 bucket is shared state keyed by $(t,k)$, so a retry after deferral meets the same depleted bucket.
This preserves fairness across activity and worker boundaries.
On a cold start, a bucket without stored state is created at full capacity inside the atomic admission call.
Concurrent first calls therefore cannot over-admit.

\subsection{Degraded-open fallback}

If the admission operation times out or the shared state is unreachable, the gate logs a warning and calls the provider unchanged.
During an outage, the system behaves exactly as before Backpressure was installed.
I3 thus degrades a failure to the naive baseline instead of losing work.
Unit tests and a local fault-injection smoke test verify this path locally:
with the shared state stopped mid-run, calls proceed ungated and admission resumes on restore.

\section{Evaluation}
\label{sec:evaluation}

We run five experiments.

\noindent\textbf{E1 (Retry waste).} We measure how much upstream retry waste worker-side admission removes under saturation.
\textbf{E2 (Slot occupancy).} We test whether timer-based deferral decouples slot occupancy from provider backoff.
\textbf{E3 (Priority order).} We test whether the gate preserves the configured priority order under contention.
\textbf{E4 (Fair share).} We test whether per-tenant service tracks the configured weighted fair share.
\textbf{E5 (Token binding).} We test whether these properties persist when tokens, not requests, bind.

\subsection{Testbed and configurations}
\label{sec:eval:testbed}

All experiments run on a fixed Google Cloud topology in one zone (\texttt{europe-west4-a}).
An \texttt{e2-standard-4} node runs the worker, and two \texttt{e2-medium} nodes run the mock upstream and Valkey. %
Temporal uses an \texttt{e2-standard-2} PostgreSQL host and three control VMs (frontend, history, two matching shards).
Valkey is deployed for every run so that all configurations see identical infrastructure.
The mock upstream enforces a fixed-window cap of \qty{60}{\request\per\second}, plus \qty{12000}{\token\per\second} on the multi-resource ladder.
It returns \texttt{HTTP 429} above either bucket and simulates \qtyrange{50}{150}{\milli\second} of service latency.
These per-second windows compress the providers' per-minute caps and preserve the request-to-token ratio.

\noindent\textbf{Fidelity of the mock upstream.}
The mock serves an OpenAI-shaped endpoint and reproduces the denial signal (rate-limit headers, a provider-shaped error body, \texttt{Retry-After} in milliseconds).
A call passes only if both the request and token limiters admit it, so either dimension can bind.
A live endpoint differs in three ways.
Token cost is estimated from request size rather than a tokenizer, and there are no per-model sub-quotas.
Rejected requests do not count against the window, so measured baseline retry waste is a lower bound.
We run a fixed window (the limiter also supports sliding windows and continuously refilling buckets) because it is the harder case.
Provider capacity arrives in steps while the gate spends it smoothly.
The window model never enters the gate's parameters, which derive from the rate $C_d$ alone (A1), and all configurations meet the same upstream.

We compare four configurations.
\sNaive{} uses SDK retry with exponential backoff on 429 (industry baseline).
\sTn{} (temporal-native) sets Temporal's matching-service \texttt{queue-rps-limit=60} plus per-workflow \texttt{Priority} and \texttt{fairness\_key} (strongest stock baseline).
\sLitellm{} is the open-source LiteLLM proxy in front of the same upstream, with client-side limiting.
Each tenant's virtual key gets a static weight-proportional cap slice per \qty{60}{\second} window.
The slices total \num{3600} RPM and \num{720000} TPM and sum to the cap without oversubscription.
\sBp{} is our gate with priority weights $\omega=(0.5,0.3,0.2)$.
We run each (workload, configuration) pair five times.
We report means, and figures add \qty{95}{\percent} Student-t confidence intervals.
The fairness metric is achievement-Jain over the active queues $\mathcal{Q}$: $J = (\sum_q r_q)^2 / (|\mathcal{Q}| \sum_q r_q^2)$.
Jain's index measures how evenly values are distributed.
It is 1 when all queues attain the same value and minimal ($1/|\mathcal{Q}|$) when one queue captures everything~\cite{jain1998quantitativemeasurefairnessdiscrimination}.
A queue's achievement ratio $r_q = \text{observed}_q/\text{ideal}_q$ is observed service over its WFQ (one dimension) or DRF (several) entitlement.
Thus $J=1$ means every queue attains the same fraction of its entitled share, not equal raw throughput.

\begin{table}[t]
    \centering
    \scriptsize
    \caption{Evaluation workloads.}
    \label{tab:workloads}
    \begin{tabular}{@{}l>{\raggedright\arraybackslash}p{6.1cm}@{}}
        \toprule
        \textbf{Workload} & \textbf{Description} \\
        \midrule
        \multicolumn{2}{@{}l}{\textit{Single resource (\qty{60}{\request\per\second} cap, 5 repeats)}} \\
        uncongested & one tenant at \qty{25}{\percent} of the cap, no contention \\
        overload & one tenant saturating the cap, isolates retry waste \\
        cross-priority & six tenants, three classes, concurrency 4/8/16 for P1/P2/P3 \\
        \addlinespace[2pt]
        \multicolumn{2}{@{}l}{\textit{Multi resource (adds \qty{12000}{\token\per\second}, 5 repeats)}} \\
        request-bound & 100-token prompts, the RPM ceiling binds \\
        token-bound & 400-token prompts, the TPM ceiling binds \\
        mixed-dominant & 1$\times$800- and 2$\times$80-token tenants, dominant resources differ \\
        noisy-estimates & random output length, estimates diverge from reported usage \\
        reservation & noisy-estimates with worst-case reservation (\cref{sec:design:reconcile}) \\
        reservation-capped & reservation, completions clamped to \texttt{max\_tokens=200} \\
        \bottomrule
    \end{tabular}
\end{table}

\subsection{Single-resource workloads}
\label{sec:eval:single}

The single-resource workloads (\cref{tab:workloads}) isolate one effect each.
We report cross-priority in full because it maximizes the fairness gap.
Its two tenants per class run 4/8/16 workflows for P1/P2/P3.
Any slot-proportional dispatch (\sNaive{}, \sTn{}) thus splits the cap $4\!:\!8\!:\!16$ and inverts priority, with P3 served at $\approx 4\times$ P1's rate.

\begin{figure}[t]
    \centering
    \includegraphics[width=\columnwidth]{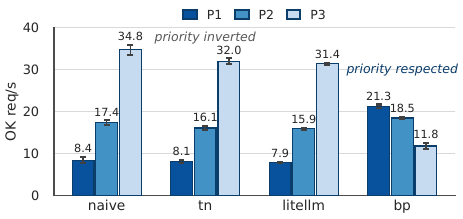}
    \caption{Cross-priority OK throughput per class (contention window, \qty{95}{\percent} CIs, 5 repeats).}
    \label{fig:w9-fairness}
\end{figure}

Backpressure restores the configured staircase in successful RPS (\cref{fig:w9-fairness}: P1 $=21.3$, P2 $=18.5$, P3 $=11.8$).
It lifts achievement-Jain from 0.60 to 0.89 and reduces P1 HTTP $p_{95}$ about \qty{\sim 8}{\times} (\qty{3506}{\milli\second} $\to$ \qty{415}{\milli\second}).
Mean worker-slot occupancy per call drops \qty{63}{\percent} (\qty{709}{\milli\second} $\to$ \qty{259}{\milli\second}).
Deferral returns the slot during the wait instead of holding it through a retry chain.
Mean OK RPS is \qty{\sim 10}{\percent} lower (55.5 vs 49.6), as the same work spreads over a longer schedule.
This answers E1, as the 429 share falls \qty{\sim 6}{\times} versus \sNaive{} (\qty{23.2}{\percent} $\to$ \qty{4.1}{\percent}), and E2, as slot occupancy decouples from provider backoff.
\Cref{fig:w9-fairness} shows \sBp{} serving P1\,$>$\,P2\,$>$\,P3, while the baselines invert the order toward the cap-blind $1/|\mathcal{T}|$ allocation.
This answers E3 (the configured ordering holds) and E4 (per-tenant service tracks the WFQ ideal).
\sLitellm{} holds the lowest 429 share (\qty{1.2}{\percent}) through client-side limiting but stays priority-blind.
P1 receives 7.9 OK RPS, the same as \sNaive{}, mean throughput falls to 35.4 RPS, and achievement-Jain (0.67) sits far below \sBp{}.
\sTn{} reaches \qty{3.1}{\percent} 429s by pinning dispatch, which leaves no backlog to reorder, so its P1 tail is the worst (\qty{3692}{\milli\second}).
Overload and uncongested runs corroborate this.
On overload (single tenant, no fairness confound), \sBp{} cuts the 429 share from 14.2 to \qty{6.0}{\percent}.
It also cuts slot occupancy from 412 to \qty{224}{\milli\second} at the highest throughput of all.
In the uncongested run, all four configurations tie within noise on latency and throughput at zero 429s.

\subsection{Multi-resource workloads}
\label{sec:eval:multi}

The single-resource results show that the gate works when requests bind.
The multi-resource workloads add a \qty{12000}{\token\per\second} cap and test whether this still holds when tokens bind (\cref{tab:multidim}).

\begin{table*}[t]
    \centering
    \scriptsize
    \caption{Multi-resource results overview (mean of 5 repeats, best in bold). 429 shares carry \qty{95}{\percent} CIs of at most 1.9\,pp. DRF Jain uses the pooled contention window.}
    \label{tab:multidim}
    \begin{tabular}{@{}llrrrrrrrr@{}}
        \toprule
        & & \multicolumn{4}{c}{\bfseries HTTP 429 share (\%) $\downarrow$} & \multicolumn{4}{c}{\bfseries DRF Jain (observed/ideal) $\uparrow$} \\
        \cmidrule(lr){3-6}\cmidrule(lr){7-10}
        \textbf{Workload} & \textbf{Binds} & \sNaive & \sTn & \sLitellm & \sBp & \sNaive & \sTn & \sLitellm & \sBp \\
        \midrule
        request-bound         & RPM   & 22.6 & 2.9          & \textbf{0.0} & 3.1  & 0.888 & 0.883 & 0.887 & \textbf{0.990} \\
        token-bound           & TPM   & 41.3 & 41.1         & 17.2 & \textbf{6.1} & 0.889 & 0.888 & 0.888 & \textbf{0.995} \\
        mixed-dominant        & mixed & 32.8 & 24.2         & 9.5  & \textbf{6.7} & 0.722 & 0.753 & 0.694 & \textbf{0.996} \\
        noisy-estimates       & TPM   & 52.6 & 52.5         & 34.7 & \textbf{6.0} & 0.890 & 0.885 & 0.889 & \textbf{0.986} \\
        reservation           & TPM   & 52.4 & 52.0         & 34.6 & \textbf{3.6} & 0.886 & 0.883 & 0.892 & \textbf{0.996} \\
        reservation-capped    & TPM   & 52.9 & 52.1         & 29.2 & \textbf{3.6} & 0.881 & 0.893 & 0.885 & \textbf{0.998} \\
        \bottomrule
    \end{tabular}
\end{table*}

\begin{figure*}[t]
    \centering
    \includegraphics[width=0.46\textwidth,trim=0 0 0 19pt,clip]{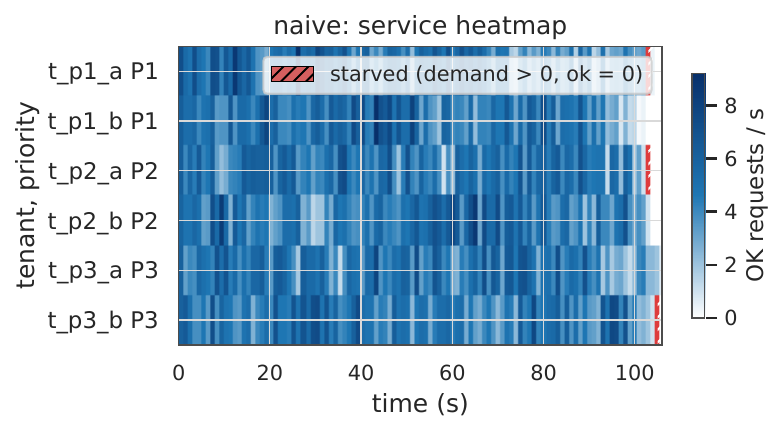}
    \hfill
    \includegraphics[width=0.46\textwidth,trim=0 0 0 19pt,clip]{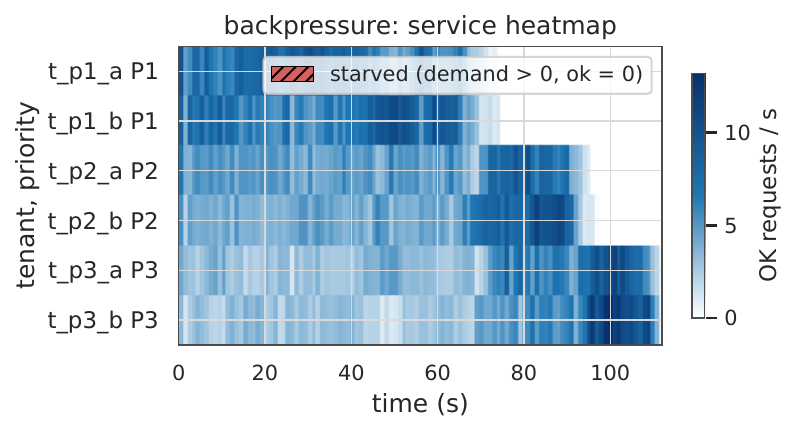}
    \caption{Token-bound service heatmaps. Rows are tenants with P1 on top, blue is served OK req/s, and hatched red marks starvation. Left: \sNaive{} serves every priority uniformly. Right: Backpressure drains P1 first and P3 last, a priority staircase although tokens, not requests, bind.}
    \label{fig:w13-heatmap}
\end{figure*}

\paragraph{Token-bound workload}
When 400-token prompts make TPM bind at \qty{\approx 30}{\request\per\second}, \sTn{}'s \texttt{queue-rps-limit=60} never engages.
\sTn{} thus floods the upstream exactly like \sNaive{} (\qty{41.1}{\percent} vs \qty{41.3}{\percent} 429s).
Retuning it to the token-implied rate ($\approx 30$) would hold only for this prompt size, and no static request ceiling suffices (mixed-dominant).
\sLitellm{}'s client-side limiter roughly halves the waste (\qty{17.2}{\percent}).
\sBp{} instead measures tokens directly and cuts it to \qty{6.1}{\percent}.
Only \sBp{} produces a priority staircase under token binding.
Its OK RPS is $14.3/8.8/5.9$ (P1/P2/P3) versus a flat $\approx 10$ for all baselines.
Its mean workflow completion is \qty{57.0}{\second} for P1 against \qty{104.7}{\second} for P3, while the baselines stay flat (\cref{fig:w13-heatmap}).
DRF Jain is $\approx 0.89$ for all baselines (within noise), 0.995 for \sBp{}.
Worker in-activity stall (\texttt{act\_succeed} mean) is $\approx 8\times$ lower under \sBp{} (\qty{211}{\milli\second}) than under \sNaive{}/\sTn{} (1634/\qty{1677}{\milli\second}), with \sLitellm{} at \qty{1452}{\milli\second}.

\paragraph{Mixed-dominant workload}
Of three equal-weight P1 tenants, one is token-heavy (TPM-dominant) and two are token-light (RPM-dominant).
Both caps bind: under \sBp{}, \qty{46}{\percent} of the 429s hit the request cap and \qty{54}{\percent} the token cap.
\sNaive{} and \sTn{} let the heavy take $1.76\times$ its fair share while the light tenants get about half of theirs ($\approx 0.52/0.55$), because equal request counts do not mean equal dominant shares.
\sLitellm{}'s static ceilings do not help (heavy $1.73\times$, light $0.47$), and its DRF Jain ($0.69$) is the worst.
\sBp{} limits the heavy tenant to $0.99$ and lifts the light ones to $\approx 0.80$.
DRF Jain rises from $0.72/0.75$ to $0.996$, which none of the baselines achieve.

\paragraph{Noisy estimates and reservation}
With random completion lengths, admission on input-based estimates alone yields \qty{39.0}{\percent} 429s.
Reconciliation (\cref{sec:design:reconcile}) cuts this to \qty{6.0}{\percent} at \qty{\approx 17}{\request\per\second} OK throughput and \qty{95}{\percent} of the token cap.
P1 HTTP $p_{95}$ falls to \qty{7.7}{\second}, against \qty{14.5}{\second} under \sNaive{}.
Worst-case reservation holds \qty{3.6}{\percent} 429s at \qty{94}{\percent} of cap, against \qty{74}{\percent} without the refund.
Clamping completions to \texttt{max\_tokens} reaches \qty{96}{\percent} of cap at \qty{3.6}{\percent} 429 share, but changes the offered work (not comparable).
This answers E5: priority ordering and DRF persist whichever dimension binds.

\subsection{Ablation: attributing the gains}
\label{sec:eval:ablation}

\begin{table}[t]
    \centering
    \footnotesize
    \caption{Token-bound ablation (5 repeats).}
    \label{tab:ablation}
    \begin{tabular}{@{}lrr@{}}
        \toprule
        \textbf{Configuration} & \textbf{429 share (\%)} & \textbf{DRF Jain} \\
        \midrule
        \sNaive{} (no gate) & 41.6 & 0.878 \\
        \sBp{} without DRF      & 38.4 & 0.959 \\
        \sBp{} without priority & 3.2  & 0.886 \\
        \sBp{} (full)       & 5.8  & \textbf{0.997} \\
        \bottomrule
    \end{tabular}
\end{table}

We disable Backpressure's mechanisms one at a time (\cref{tab:ablation}).
Gating itself (\cref{sec:timer}) removes most of the retry waste.
On token-bound, every gated arm cuts \sNaive{}'s \qty{41.6}{\percent} 429 share to single digits, except \sBp{} without DRF, which cannot see tokens.
DRF has the largest effect on waste when tokens bind.
Without it the share jumps back to \qty{38.4}{\percent}, close to \sNaive{}, against \qty{5.8}{\percent} with it.
Its effect on DRF Jain is smaller (0.959 to 0.997) because all tenants here send the same 400-token prompt.
Equal requests then already approximate equal dominant share, so a token-blind gate still allocates fairly and merely wastes the cap.
The fairness effect appears once dominant resources differ (mixed-dominant).
Priority shaping supplies the ordering.
On cross-priority, \sBp{} without priority still reaches achievement-Jain 0.839 (full: 0.889), since tenant-level fair share equalizes the six queues.
The index, however, ignores ordering, and only the full gate yields P1$>$P2$>$P3.
Both layers are needed: DRF handles the binding dimension, and priority shaping handles the class order.

\subsection{Gate overhead}
\label{sec:eval:overhead}

The gate adds one Valkey round trip per call.
In the uncongested run, the cap never binds, so the gate never defers and its overhead is within noise.
Worker in-activity time (\qty{166}{\milli\second}) matches \sNaive{}'s (\qty{172}{\milli\second}), and median HTTP latency is no worse (184 vs \qty{201}{\milli\second}).
\sLitellm{}'s client-side limiter instead adds \qty{\approx 50}{\percent} in-activity stall on every call (\qty{252}{\milli\second}).
Under contention, the gate imposes a wait, but as a durable timer with the slot released, not as held slot time.
This deferral ($p_{95}$ \qty{1.9}{\second} on cross-priority) drives the \qty{63}{\percent} slot-occupancy reduction.
Across \num{88000} evaluations the Lua step takes 1.5/4.9/\qty{9.0}{\milli\second} at the median/$p_{95}$/$p_{99}$.
At these rates, the primary stays far below its throughput ceiling (measuring it is future work).

\subsection{Trade-offs}
\label{sec:eval:tradeoffs}

Priority shaping raises the median of deprioritized classes under contention (P3 on cross-priority).
Under mild load, \sBp{}'s median matches \sNaive{}'s.
Shaping also lengthens the drain span by up to $\approx$\,\qty{12}{\percent}, the same trade an operator opts into by defining priority classes.
The work-conserving L2 leaks $\approx$ \qtyrange{10}{15}{\percent} across priorities under joint saturation, so strict-priority installations would need a hard filter.
Backpressure redistributes the cap rather than raising throughput, and its one Valkey primary serializes every admission, so scaling past it needs sharding by target.

\section{Related Work}
\label{sec:related}

We group related work by who owns the bottleneck.

\paragraph{Admission control and overload management}
The Staged Event-Driven Architecture~\cite{welsh2001seda} and Breakwater~\cite{cho2020breakwater} decompose services into stages with per-stage admission and closed-loop feedback on locally observed queuing delay.
Both assume the controller owns the protected resource.
Backpressure extends this line to a worker-side gate before a third-party endpoint whose only congestion signal is a post-hoc 429.

\paragraph{Fair queuing and distributed rate limiting}
Our L3 allocator builds on fair queuing.
WFQ~\cite{demers1989wfq} and DRR~\cite{shreedhar1995drr} approximate Generalized Processor Sharing~\cite{parekh1993gps}, an idealized fluid server, at low cost.
DRF~\cite{ghodsi2011drf} extends the single-resource objective to multiple resources, for a datacenter scheduler that owns and observes them all.
Distributed rate limiters (Raghavan et~al.~\cite{drl}, Seawall~\cite{seawall}) likewise assume the enforcing system controls the limited resource and observes queue depths.
Industry global rate limiters (Google's Doorman~\cite{doorman_design}, Salesforce's coordinated limiting~\cite{dedeo2019salesforce}) coordinate clients against a limit the enforcing system owns.
Backpressure's shared admission state instead enforces an externally owned cap that it cannot observe, using configured limits.

\paragraph{Backpressure in dataflow systems}
The term \emph{backpressure} originates in streaming dataflow, where a slow consumer signals the producer to slow down.
Reactive Streams~\cite{reactive_streams_spec} and Flink's credit-based flow control~\cite{flink_network_stack} implement it end to end.
The producer can be slowed because it sits inside the same system.
Our setting inverts that assumption, as the bottleneck is a third-party API the operator cannot push back on.
Backpressure therefore applies the pressure inward, onto the worker, rather than blocking the source.
Serverless dataflow runtimes such as Cloudburst~\cite{sreekanti2020cloudburst} and Kappa~\cite{zhang2020kappa} cannot push back on external calls either, and neither addresses fair admission against a shared cap.

\paragraph{Multi-tenant fairness in LLM serving}
A large body of work optimizes inference inside a fabric the operator owns.
PagedAttention~\cite{kwon2023pagedattention} raises throughput through memory management and batching, and the Virtual Token Counter (VTC)~\cite{sheng2024vtc} adds fairness to that scheduler. %
All assume batch, token-count, and memory visibility inside the engine.
Backpressure addresses the strictly upstream problem: fair admission before work reaches such a fabric, most commonly an unmodifiable third-party API.

\paragraph{Workflow engines and HTTP retry gateways}
Engine-internal advances~\cite{zhuang2023exoflow,burckhardt2022netherite,jia2021boki} do not change where admission against an external cap must run.
HTTP retry gateways~\cite{litellm} centralize routing and retries at a layer that cannot release a worker slot.
These gateways compose cleanly with Backpressure for routing and key management.

\section{Discussion and Limitations}
\label{sec:discussion}

With Backpressure, a cap the operator cannot change can still be scheduled fairly, as the waiting happens outside the worker slot.
Neither the fairness rule nor the deferral depends on how admission state is stored or which engine hosts the wait.
The caps must be static and known (A1 to A3), which bounds where the approach applies.
Adapting to provider-reported or drifting limits is not covered.
Client-side probing in the Additive Increase Multiplicative Decrease (AIMD) tradition~\cite{aimd} is a natural starting point.
We evaluate only on Temporal, but other engines should behave similarly, as the approach needs only a durable sleep, which most of them offer~\cite{stepfunctions,durable_functions,airflow}.
The synthetic upstream reproduces published 429 semantics but not every detail of a live endpoint, which we argue suffices to evaluate our prototype.
Admission is serialized by one authoritative copy of the shared state.
We do not benchmark its ceiling or sharding by target; at the measured admission latency, one primary serializes far more admissions per second than current provider caps allow.

\section{Conclusion}
\label{sec:conclusion}
Workflow engines assume elastic downstream services, and none of them links an external cap to the work it dispatches.
LLM providers make that gap a daily operational problem.
Their caps are hard, owned by a third party, and binding on requests and tokens at once.
Backpressure adds the link at the worker, the one place the operator still controls.
Our gate lifts achievement-Jain from 0.60 to 0.89 on a cross-priority workload and cuts high-priority tail latency roughly eightfold.
When tokens bind, it holds dominant-resource fairness near 0.99.
As more applications depend on scarce third-party APIs, we believe workflow engines should offer such admission control as a built-in feature.

\section*{Acknowledgement of Generative AI Use}
Generative AI assistants (Anthropic Claude, OpenAI ChatGPT) supported an initial broad survey of related work and provided reviews of drafts.
All citations were read and verified by the authors, who take full responsibility for all content.

\balance

\bibliographystyle{IEEEtran}
\bibliography{bibliography}

\end{document}